\documentclass[useAMS,usenatbib,usegraphicx]{mn2e}

\def\cs{$\chi^2$}
\def\er{Equation~\ref}
\def\fr{Figure~\ref}
\def\kmps{km\,s$^{-1}$}

\def\mgtwo{Mg{\sc \,ii}}

\def\scr{Section~\ref}
\def\spose#1{\hbox to 0pt{#1\hss}} 
\def\tr{Table~\ref}
\def\approxlt{\mathrel{\spose{\lower 3pt\hbox{$\sim$}}
        \raise 2.0pt\hbox{$<$}}}

\def\approxgt{\mathrel{\spose{\lower 3pt\hbox{$\sim$}}
        \raise 2.0pt\hbox{$>$}}}

\def\altwo{Al{\sc \,ii}}

\def\approxgt{\mathrel{\spose{\lower 3pt\hbox{$\sim$}}
        \raise 2.0pt\hbox{$>$}}}
\def\approxlt{\mathrel{\spose{\lower 3pt\hbox{$\sim$}}
        \raise 2.0pt\hbox{$<$}}}

\def\catwo{Ca{\sc \,ii}}

\def\cone{C{\sc \,i}}

\def\crtwo{Cr{\sc \,ii}}
\def\cs{$\chi^2$}
\def\csn{$\chi^2/\nu$}
\def\daa{$\Delta \alpha/\alpha$}
\def\dmm{$\Delta \mu/\mu$}
\def\dmmavtw{$\langle\Delta \mu/\mu\rangle^{\rm weighted}_{\rm total}$}
\def\dxx{$\Delta x/x$}

\def\dxxavq{$\langle\Delta x/x\rangle_{\rm abs}$}

\def\dxxavtw{$\langle\Delta x/x\rangle^{\rm weighted}_{\rm total}$}

\def\er{Equation~\ref}
\def\fetwo{Fe{\sc \,ii}}
\def\fr{Fig.~\ref}
\def\kmps{km s$^{-1}$}

\def\mgone{Mg{\sc \,i}}
\def\mgtwo{Mg{\sc \,ii}}
\def\mntwo{Mn{\sc \,ii}}

\def\nitwo{Ni{\sc \,ii}}

\def\sitwo{Si{\sc \,ii}}	
\def\spose#1{\hbox to 0pt{#1\hss}} 
\def\stwo{S{\sc \,ii}}

\def\titwo{Ti{\sc \,ii}}

\def\zem{$z_{\rm em}$}
\def\zntwo{Zn{\sc \,ii}}

\def\zuv{$z_{\rm UV}$}
\def\zavuv{$\langle z_{\rm UV} \rangle$}
\def\zrad{$z_{\rm 21}$}

\def\znf{Q0952$+$179}
\def\oot{Q1127$-$145}
\def\ott{Q1229$-$021}
\def\ztt{Q0235$+$164}
\def\zet{Q0827$+$243}
\def\otto{Q1331$+$170}
\def\oof{Q1157$+$014}
\def\zff{Q0458$-$020}
\def\zfe{Q0438$-$436}

\def\aj{AJ}%
\def\apj{ApJ}%
\def\apss{Ap\&SS}%
\def\aap{A\&A}%
\def\mnras{MNRAS}%
\def\prd{Phys.~Rev.~D}%
\def\prl{Phys.~Rev.~Lett.}%
\def\rmp{Rev. Mod. Phys.}
\def\nat{Nature}%
\def\aplett{Astrophys.~Lett.}%
\title[Radio and optical 
quasar absorption lines]{Probing
variations in fundamental constants with radio and
optical quasar absorption-line observations}

\author[P. Tzanavaris et al.] 
{P. Tzanavaris$^{1}$\thanks{E-mail:pana@astro.noa.gr (PT)}\footnotemark[1]\thanks{Present
address:Institute of Astronomy and Astrophysics, National Observatory
of Athens, I. Metaxa \& V. Pavlou, 152 36 Penteli, Greece},
M. T. Murphy$^{2}$, J. K. Webb$^{1}$, V. V. Flambaum$^{1,3}$,
S. J. Curran$^{1}$\\ 
$^{1}$School of Physics, The University of New
South Wales, NSW 2052, Sydney, Australia\\ 
$^{2}$Institute of
Astronomy, Madingley Road, Cambridge CB3 0HA\\
$^{3}$Physics Division, Argonne National Laboratory, Argonne, IL 60439, USA.}

\begin{document}
\date{Accepted 2006 October 09. Received 2006 October 02; in original form
2006 August 25}

\pagerange{\pageref{firstpage}--\pageref{lastpage}} \pubyear{2005}

\maketitle

\label{firstpage}

\begin{abstract}
  Nine quasar absorption spectra at 21-cm and UV rest-wavelengths are
  used to estimate possible variations in $x\equiv \alpha^2 g_{\rm p} \mu$,
  where $\alpha$ is the fine structure constant, $g_{\rm p}$ the proton
  $g$-factor and $\mu\equiv m_{\rm e}/m_{\rm p}$ is the
  electron-to-proton mass ratio.  We find
  \dxxavtw~$=(0.63\pm0.99)\times10^{-5}$ over a redshift range
  $0.23\approxlt z_{\rm abs}\approxlt 2.35$ which corresponds to
  look-back times of 2.7--10.5 billion years. A linear fit against
  look-back time, tied to \dxx\ $=0$ at $z=0$, gives a best-fit rate
  of change of $\dot{x}/x=(-0.6\pm1.2)\times 10^{-15}{\rm\, yr}^{-1}$.
  We find no evidence for strong angular variations in $x$ across the
  sky. Our sample is much larger than most previous samples and
  demonstrates that intrinsic line-of-sight velocity differences
  between the 21-cm and UV absorption redshifts, which have a random
  sign and magnitude in each absorption system, limit our precision.
  The data directly imply that the average magnitude of this
  difference is $\Delta v_{\rm los}\sim 6$~\kmps.

  Combining our \dxx\ measurement with absorption-line constraints on
  $\alpha$-variation yields strong limits on the variation of $\mu$.
  Our most conservative estimate, obtained by assuming no variations
  in $\alpha$ or $g_{\rm p}$ is simply \dmm\ $=$ \dxxavtw.  If we use
  only the four high-redshift absorbers in our sample, we obtain
  \dmm~$=(0.58\pm1.95)\times 10^{-5}$, which agrees ($2\sigma$) with
  recent, more direct estimates from two absorption systems containing
  molecular hydrogen, also at high redshift, and which have hinted at
  a possible $\mu$-variation, $\Delta\mu/\mu = (-2.0\pm0.6)\times
  10^{-5}$.  Our method of constraining \dmm\ is completely
  independent from the molecular hydrogen observations.  If we include
  the low-redshift systems, our \dmm\ result differs significantly from
  the high-redshift molecular hydrogen results.  We detect a
  dipole variation in $\mu$ across the sky, but given the sparse
  angular distribution of quasar sight-lines we find that this model
  is required by the data at only the 88 per cent confidence
  level. Clearly, much larger samples of 21-cm and molecular hydrogen
  absorbers are required to adequately resolve the issue of the
  variation of $\mu$ and $x$.
\end{abstract}


\section{Introduction}
The existence of extra spatial dimensions remains an open question in
theoretical physics. The answer to this question is closely linked to
attempts at super-unification, which has become a modern holy grail.
In turn, the existence of extra spatial or temporal dimensions may be
inferred by the detection of spatial or temporal variations in the
values of coupling constants, such as the fine structure constant,
$\alpha$ \citep[see][for reviews]{2003UzanRMP,2005HAM...KFP}.  This
constant provides a measure of the strength of the electromagnetic
interaction and is truly fundamental because it is dimensionless and
thus independent of any system of units. Other such constants are the
electron-to-proton mass ratio, $\mu$, the proton gyromagnetic factor
$g_{\rm p}$, as well as the constant $x$ defined as
\begin{equation}
x\equiv \alpha^2 g_{\rm p} \mu \ ,
\end{equation}
which combines all three.  Apart from the question of extra
dimensions, any confirmed detection of a variation in the value of any
of these `constants' would in itself be a ground-breaking
result.  For example, in some popular models this variation is related
to violation of the Principle of Equivalence.

There have been several attempts to constrain the variation of
fundamental constants either experimentally or observationally. Atomic
clocks \citep[see
e.g.][]{1995PRL...74...3511P,2001PhysScript...95...50S,2003PRL...90...150801M,2004Flambaum,2006FlamTed}
and the Oklo nuclear reactor
\citep{1976Nat...264...340S,2000NPB...573...377F,
  2002PRD...65...103503F,2002PRD...66...045022O,2004...PRD...69...121701L}
provide local, i.e. earthbound and relatively recent, $z\sim 0$,
constraints.  The cosmic microwave background provides constraints at
$z\sim 1000$
\citep{2001PRD...63...083505L,2001PRD...64...103505A,2002PRD...66...023505M}.
Further constraints may be obtained from Big Bang Nucleosynthesis data
\citep[see e.g.][]{2004PhRvD..69f3506D}.

However, one very promising way to detect such variations is
astronomical spectroscopy of gas clouds which intersect the lines of
sight to distant quasars.  This technique allows one to obtain
constraints essentially over all of the intermediate redshift range,
which is crucial for probing variations with cosmic time.
Additionally, results achieve very high precision, due to the advent
of 10-m class telescopes equipped with state-of-the art
high-resolution spectrographs, in particular Keck/HIRES in Hawaii and
VLT/UVES in Chile.
 
In this type of work, the gas clouds probed are often damped Lyman
$\alpha$ systems (DLAs). These systems have high neutral hydrogen
column densities ($N_{\rm H I} \approxgt 10^{20} \ {\rm cm}^{-2}$) at
precisely determined redshifts, thus providing excellent probes of the
bulk neutral atomic hydrogen in the distant Universe. Furthermore,
DLAs contain many different heavy elements which absorb in the
rest-frame ultraviolet (for atomic species) or millimetre (for
molecular species), giving rise to a multitude of absorption lines.
With the latest generation of optical telescopes rest-frame
ultraviolet lines can be observed with resolutions of the order of
6~\kmps\ so that redshifts can be determined to six significant
digits.  This is crucial as all methods using this type of data aim to
detect small shifts induced by a possible variation of constants which
are of different magnitude for different transitions.

Using such data, \citet{1999PhRvL..82..888D} and
\citet{1999PhRvL..82..884W} developed the highly sensitive {\it
  many-multiplet method} and in a series of papers applied it to
rest-frame ultraviolet {\it atomic} quasar absorption lines, obtained
with Keck/HIRES. Their work provided evidence that $\alpha$ is smaller
at redshifts $0.2 < z < 3.7$ at the fractional level of $\Delta
\alpha/\alpha=(-0.57\pm0.10) \times 10^{-5}$ \citep{1999PhRvL..82..884W,2001-A-MNRAS.327.1208M, 2001PhRvL..87i1301W,2003ApSS.283..565W,2003MNRAS.345..609M,2004LNP...648..131M}.  After
thorough investigation \citet{2001-B-MNRAS.327.1223M} and 
  \citet{2003ApSS.283..577M} ruled out a large range of known systematic
effects.  However the issue remains controversial after another group
found no evidence for a change in $\alpha$ using VLT/UVES spectra
\citep[$\Delta \alpha/\alpha=(-0.06\pm0.06) \times 10^{-5}$ for $0.4<z<2.3$,]
[]{2004AA...417..853C,2004PhRvL..92l1302S}.

Strong constraints on variations in the electron-to-proton mass ratio,
$\mu$, can also be obtained from optical quasar absorption studies
\citep{1975ApL....16....3T}. The first confirmed {\it molecular} hydrogen
quasar absorption line detections were obtained by 
\citet{1985MNRAS.212..517L}, and confirmed by \citet{1988ApJ...324..267F}.
However, 
although the large number of H$_{2}$ transitions 
in the Lyman and Werner bands have different dependencies on $\mu$,
making them useful probes of $\mu$-variation, unfortunately the
relatively small fraction of DLAs in which H$_2$ is detected has
limited this work so far
\citep{1993JETPL.....58..237V,
  2002AstL...28..423I,2004PRL...92...101302U,2005A&A...440...45I}.
Nevertheless, the recent study of two high signal-to-noise spectra has
yielded hints that $\mu$ may have been smaller at redshifts between
$\sim$2.5--3.1 \citep{2006Reinhold} \footnote{Note that
  \citet{2006Reinhold} use the inverse definition of $\mu$
  (i.e.~$\mu\equiv m_{\rm p}/m_{\rm e}$) compared with this paper.}.

A potentially more sensitive approach is possible when, as well as
rest-frame UV, rest-frame 21-cm absorption from cold neutral hydrogen
is also detected. In such cases, variability of the constant $x$,
defined above, can be investigated. Because the ratio of 21-cm to UV
transition frequencies is proportional to $x$, it can be shown that,
if both UV and 21-cm absorption occurs at the same physical location,
the relative change in the value of $x$ between its value at redshift $z$ 
and its value in a terrestrial lab
is related to the observed absorption redshifts for rest-frame
21-cm and UV, \zrad\ and \zuv, according to
\begin{equation}\label{equ:dxx}
\Delta x/x \equiv \frac{x_z-x_0}{x_0}
= \frac{z_{\rm UV}-z_{\rm 21} } {1+z_{\rm 21}}
\end{equation}
\citep[e.g.][]{1980ApJ...236L.105T}. In addition, because $\mu$ enters
in \er{equ:dxx} via the definition of $x$, results on the variation of
$\mu$ may also be obtained, allowing completely independent tests of
\dmm\ results obtained more directly from molecular absorption lines.
However, application of this method is limited by the fact that there
are only $\sim$18 DLAs for which absorption has been observed both in
the optical and in the radio \citep[][contains a detailed
list]{2005MNRAS.356.1509C}.  Consequently, there is only a handful of
results based on this method
\citep{1979AJ.....84..699W,1980ApJ...236L.105T,
  1995ApJ...453..596C,1981ApJ...248..460W}. Among these, a
high-resolution spectrum from a 10-m class telescope is used only once
and for a single object \citep{1995ApJ...453..596C}.  Furthermore,
there has been no attempt to apply this method to a sample consisting
of more than one object. This is important for two reasons:
\begin{enumerate}
\item Any systematic effects due to spatial non-coincidence of
  hydrogen and heavy elements can only be detected if a sample of
  objects is used.
\item To detect any space-time variation of $x$, one needs several
  observations, covering a range in redshift and direction.
\end{enumerate}

The first results from such an approach were briefly presented in
\citet{2005PhRvL..95d1301T}.  In this paper we explain in detail how
these results were obtained and extend our original sample of eight
absorbers with a newly discovered 21-cm absorber. We have filled the
gap in the literature by applying this method to a sample comprising
{\it all} the best available 21-cm absorption data in conjunction with
the highest-resolution UV data available.  The sample is made up of
nine absorption systems situated along the line of sight to nine
quasars covering the absorption redshift range $\sim 0.23$ to $\sim
2.35$. The paper's structure is as follows: In \scr{sec:data} we
describe our data sample. We describe our methods and results in
\scr{sec:analysis}. Our results are discussed in \scr{sec:discussion}.

\section{Data}\label{sec:data}
\subsection{Data sample and reduction}
Details of the data used in this work are given in \tr{tab:data} and
are discussed below. Our data sample is made up of nine quasar
absorption systems along the lines of sight to nine quasars. Each
system shows both 21-cm and UV absorption (rest frame) along the line
of sight to a different quasar. Rest-frame UV absorption is observed
red-shifted in the optical region, and rest-frame 21-cm is observed
red-shifted at longer radio wavelengths. In this paper we use the
labels UV and 21-cm to distinguish the two types of observations that
we use.

Figures~\ref{fig:0952} to \ref{fig:0438} show the 21-cm and UV data
registered to a common velocity scale. We note that all data have been
corrected to the heliocentric frame.

\subsubsection{Radio data}\label{sec:rad}
The original radio 21-cm absorption spectra were not available for
this work, apart from the data for quasar \zff\ which were kindly
provided by Art Wolfe. Therefore, we digitised the 21-cm plots from
the literature (see the references in \tr{tab:data}) using the {\sc
  dexter} application \citep{2001ASPC..238..321D}, available at
NASA/ADS for on-line scanned papers or, for PDF plots, the stand-alone
version available at http://sourceforge.net/projects/dexter.

The initial step was to digitise each spectrum, obtaining
flux/abscissa values. Then a linear wavelength scale was derived for,
and assigned to, each spectrum by carefully measuring velocity or
frequency points on the x-axis of the plot.  This second step
dominates the uncertainty in the zero-point of the velocity/frequency
scale but is nevertheless not an important source of error. As an
example, we digitised a postscript plot of the \sitwo~1808~\AA\
absorption feature obtained with Keck/HIRES and compared the resulting
redshift to that obtained from the original data. The two redshifts
differed by just $1\times 10^{-6}$. A very conservative upper limit on
the absolute velocity/frequency uncertainty is given by half of the
digitised pixel size in each case.  We checked the maximum
contribution this zero-point error could make to the final estimate
for \dxx\ for each quasar and, in all cases, this was less than 25 per
cent of the statistical-plus-systematic (i.e.~increased) error bars
illustrated in \fr{fig:dxxz}. We therefore stress that the
digitisation process generally contributes a small uncertainty in
comparing the 21-cm and UV absorption redshifts.

Redshift uncertainties and details of the observations and data
reduction can be found in the references listed in \tr{tab:data}.
However, since no redshift uncertainty is given for the radio spectrum
of \oot\ in \citet{2001MNRAS.325..631K} we used the uncertainty from
\citet{2001AA...369...42K}, as these are the same observers, using the
same instrument to perform a similar type of observation.

\subsubsection{Optical (rest-frame UV) data}
Optical (rest-frame UV) high-resolution echelle spectral data for
eight of the nine quasars were taken from the European Southern
Observatory's (ESO's) archive. These were originally observed with the
UVES spectrograph on the Very Large Telescope (VLT). Data reduction
was carried out with the ESO/UVES pipeline
\citep{2000Mess...101...31B}, modified to improve the flux extraction
and the wavelength calibration precision. Post-pipeline processing was
carried out using the custom-written code {\sc uves popler}
\footnote{http://www.ast.cam.ac.uk/$\sim$mim/UVES\_popler.html}. For
each quasar, the extracted echelle orders from all exposures were
combined to form a single 1-dimensional spectrum, with cosmic ray
signatures being rejected by a sigma-clipping algorithm during the
combination. A continuum was constructed by iteratively fitting a
series of overlapping polynomials, rejecting discrepant pixels at each
iteration until convergence. For one of the quasars (\otto) we have
also included the part of the absorption spectrum for
\sitwo~1808\,\AA\ from a Keck/HIRES observation, provided by Art
Wolfe. This spectrum was reduced with the dedicated Keck/HIRES
pipeline, {\sc makee}
\footnote{http://spider.ipac.caltech.edu/staff/tab/makee/index.html},
written by Tom Barlow.

For quasar \ztt\ we digitised an absorption plot from
\citet{1992ApJ...391...48L} for a single heavy element species.

\begin{table*}
\begin{center}
  \caption{\small {\label{tab:data}}Details of our data sample. There
    is one 21-cm/UV absorption system along each quasar sight-line.
    Column 1 is the quasar name and Column 2 its emission redshift.
    For the strongest component Column 3 gives the 21-cm absorption
    redshift. The uncertainty is given in parentheses and has been
    taken from the references shown in Column 4.  The original 21-cm
    plots were also obtained from these references. However, the
    entries marked with $\dagger$ which refer only to the source of
    the 21-cm redshift uncertainty, whilst for the entry marked with
    $\ast$ there is no uncertainty in redshift in the literature (see
    text).  Column 5 gives the mean absorption redshift and standard
    deviation on the observed mean for the strongest UV component.
    Column 6 gives the UV heavy element species observed in the
    optical with the number of transitions in parentheses, if more
    than one. Column 7 gives the source for the UV data. For data
    obtained from the ESO/UVES archive, the ESO program identification
    code is given in this column and the principal investigator of the
    program is given in the footnotes.  For \otto\ we also used
    \sitwo~1808\,\AA\ Keck/HIRES data provided by A.~Wolfe.  For
    quasar \ztt\ we digitised an absorption plot from the literature
    for a single heavy element species.}
\begin{tabular}{cclccll}
\\hline
Quasar & \zem\  & \zrad\        & 21-cm         &$\langle z_{\rm UV}\rangle$& ions                                        & UV              \\\hline
\znf   &  1.478 & 0.237803(20)  & [1]           & 0.237818(6)            &\mgone, \catwo(2)                            & 69.A-0371(A)$^a$ \\
\oot   &  1.187 & 0.312656(50)  & [2]           & 0.312648(6)            &\catwo(2), \mntwo(3)                         & 67.A-0567(A)$^b$, 69.A-0371(A)$^a$\\
\ott   &  1.038 & 0.394971(4)   & [3]           & 0.395019(40)           &\catwo(2), \mntwo(3), \titwo                 & 68.A-0170(A)$^c$\\
\ztt   &  0.940 & 0.523874(100) & [4]           & 0.523829(6)            &\mgone                                       & \cite{1992ApJ...391...48L} \\
\zet   &  0.941 & 0.524757(50)  & [1]           & 0.524761(6)            &\catwo(2), \fetwo\                           & 68.A-0170(A)$^c$, 69.A-0371(A)$^a$\\
\otto  &  2.097 & 1.776427(20)  & [5]           & 1.776355(5)            &\mgone, \altwo, \sitwo, \stwo,               & 67.A-0022(A)$^d$, 68.A-0170(A)$^c$\\
       &        &               &               &                        &\cone(3), \cone$^{\ast}$, \crtwo(2), \mntwo(2),& \\
       &        &               &               &                        &\fetwo(4), \nitwo(6), \zntwo\                & \\
       &        &               &               &                        &\sitwo$^e$                                   &$\dagger$ \\ 
\oof   &  1.986 & 1.943641(10)  & [6]           & 1.943738(3)            &\mgone, \mgtwo(2), \sitwo,                   & 65.O-0063(B)$^f$, 67.A-0078(A)$^f$, \\
       &        &               &                                        &                        & \nitwo(6)                                   & 68.A-0461(A)$^g$\\
\zff   &  2.286 & 2.039395(80)  & [7]           & 2.039553(4)            &\zntwo(2), \nitwo(6), \mntwo(3),             & 66.A-0624(A)$^f$, 68.A-0600(A)$^f$, \\
       &        &               &                                        &                        &\crtwo(3)                                    & 072.A-0346(A)$^f$, 074.B-0358(A)$^h$ \\
\zfe   &  2.347 & 2.347488(12)  & [8]           & 2.347403(49)           &\zntwo(2), \mntwo, \fetwo, \sitwo            & 69.A-0051(A)$^i$, 072.A-0346(A)$^f$ \\
\hline
\end{tabular}

\end{center}
1: \cite{2001AA...369...42K};
2: \cite{2001MNRAS.325..631K}; uncertainty in redshift not available (see text);
3: \cite{1979ApJ...230L...1B};
4: \cite{1978ApJ...222..752W};
5: \cite{1979AJ.....84..699W}; A.~Wolfe, private communication: postscript plot;
6: \cite{1981ApJ...248..460W};
7: \cite{1985ApJ...294L..67W}; A.~Wolfe, private communication: original data;
8: \cite{2006MNRAS.370L..46K}.

$a$: Savaglio;
$b$: Lane;
$c$: Mall{\'e}n-Ornelas;
$d$: D'Odorico;
$e$: Same transition as in the UVES data but from Keck/HIRES;
$f$: Ledoux;
$g$: Kanekar;
$h$: Dessauges-Zavadsky;
$i$: Pettini;
$\dagger$ A.~Wolfe, private communication: original \sitwo\ data in \otto.
\end{table*}


\section{Analysis}\label{sec:analysis}
There is one absorption system for each 21-cm/UV absorption spectrum
pair in a single quasar. To apply \er{equ:dxx} we need values for
\zuv\ and \zrad.  We now explain how we chose these values.

\subsection{Assumptions}
In each absorption system we used the frequency or wavelength of
strongest absorption for each absorbing species, neutral or singly
ionised, to provide the 21-cm and UV redshifts.  We justify these
choices below but note that they need not represent the true physical
picture for our results to be useful.  Rather, they are used as a
starting point for our analysis.

\subsubsection{Strongest absorption}
It is evident from Figs.~\ref{fig:0952}--\ref{fig:0438} that in
most absorption systems there is usually complex velocity structure.
For neutral and singly ionised heavy element species there is
sufficient similarity in this structure so that corresponding
components can be identified.  Unfortunately, this is not normally the
case when one compares UV to 21-cm absorption profiles in the same
system. In other words, it is not easy to tell which 21-cm absorption
component corresponds to which UV absorption component. However, a
decision must be made because \er{equ:dxx} only holds if it is applied
to 21-cm and UV redshifts for components which are situated at the
same physical location.  It is important to realise that this
represents a fundamental limitation in the comparison of 21-cm and UV
absorption lines when deriving constraints on varying constants.
Clearly, this problem can only be overcome using a statistically large
sample of absorption systems. The method used to determine the
`centroid' of the 21-cm and UV absorption is also clearly not crucial
provided that it is unbiased and that uncertainties can be reliably
estimated.

Given the inherent dissimilarity between the 21-cm and UV absorption
profiles, many possible methods exist for choosing appropriate 21-cm
and UV redshifts. For example, for a single absorber, the various UV
profiles might be cross-correlated with the 21-cm profile or one might
attempt to identify `nearest neighbour' UV and 21-cm velocity
components. However, for the cross-correlation example, it is not
clear how to derive a meaningful uncertainty on the redshift
comparison. In the `nearest neighbour' approach, recently advocated in
\citet{2006MNRAS.370L..46K}, a detailed fit to the 21-cm and UV
profiles is required to identify all the velocity components and,
given the general variety of spectral resolutions apparent in
Figs.~\ref{fig:0952}--\ref{fig:0438}, this is difficult.  Moreover,
this approach is strongly biased towards zero $x$-variation and this
bias is again resolution-dependent. The redshift uncertainty for
individual absorbers will therefore be too large. Similarly, for a
large sample, the scatter in the values of \dxx\ will be significantly
underestimated and any additional scatter caused by intrinsic velocity
shifts between the 21-cm and UV components will be masked.  

The simplest method is to take \zrad\ and \zuv\ from the strongest
absorption component, as this is well-defined in our data.  Although
there is no general similarity in the velocity profiles, we assume
that strongest absorption occurs at the same physical location for all
species.  We stress that this is a simple, initial assumption and
later we demonstrate that we can actually use our results to test this
assumption. Given the range of options above and given the fundamental
limitation in associating 21-cm and UV components which are typically
spaced by several \kmps, it is also clear that detailed fits to the
components in order to decrease the centroid position uncertainty to
the sub-pixel level is unimportant. A cruder and simpler measurement,
such as taking the redshift of the pixel with minimum intensity, will
still involve an uncertainty which is smaller than the errors involved
in comparing the 21-cm and UV profiles.

\subsubsection{Neutral and singly ionised species}
Neutral heavy element species might be considered more likely to be
spatially coincident with 21-cm absorbing gas, which is due to neutral
hydrogen. We decided to use singly ionised species as well for the
following reasons:
\begin{enumerate}
\item In our sample, the velocity structure of neutral species was
  followed closely by singly ionised species. This was also true in
  the much larger sample in \citet{2004LNP...648..131M} used to
  constrain variations in $\alpha$.
\item Extensive work by \citet{2003ApJ...582...49P} more generally
  indicates that ionisation fraction, abundance and dust-depletion may
  not change significantly along an absorption complex.
\item In any case, we only used the simple option of strongest
  components and these, in the overwhelming majority of cases,
  corresponded very clearly between neutral and singly ionised species
  (see Figs.~\ref{fig:0952}--\ref{fig:0438}).
\end{enumerate}
By using both neutral and singly ionised species, we obtained a
significantly larger sample of UV lines for each absorber. This
allowed us to investigate systematics for individual absorbers for the
first time. Note though that there is one exception to our approach.
This concerns \catwo\ which has {\it not} been used in our analysis,
although it is singly ionised and shown in the figures for
illustration purposes only (but see \scr{sec:discussion}).  The reason
for this is that the \catwo\ ionisation potential, 11.871~eV, is such
that one would not expect this species to trace neutral components
well compared to the other singly ionised species used.  Since DLAs
have high neutral hydrogen column densities, one expects only photons
with energies $< 13.6$~eV to penetrate the radiation shielding gas
envelope.  Such photons would then be unable to ionise species such as
\fetwo\ (ionisation potential 16.18~eV) or \mntwo\ (ionisation
potential 15.64~eV), but they would still be able to ionise \catwo.
Thus all singly ionised species used, apart from \catwo, are much more
likely to be the dominant ionised species in an absorption system.
For this reason we did not use this transition, just as we did not use
higher ionisation species.

\subsection{Redshift determination}
\begin{figure*}
\includegraphics[width=1.5\columnwidth, angle=0]{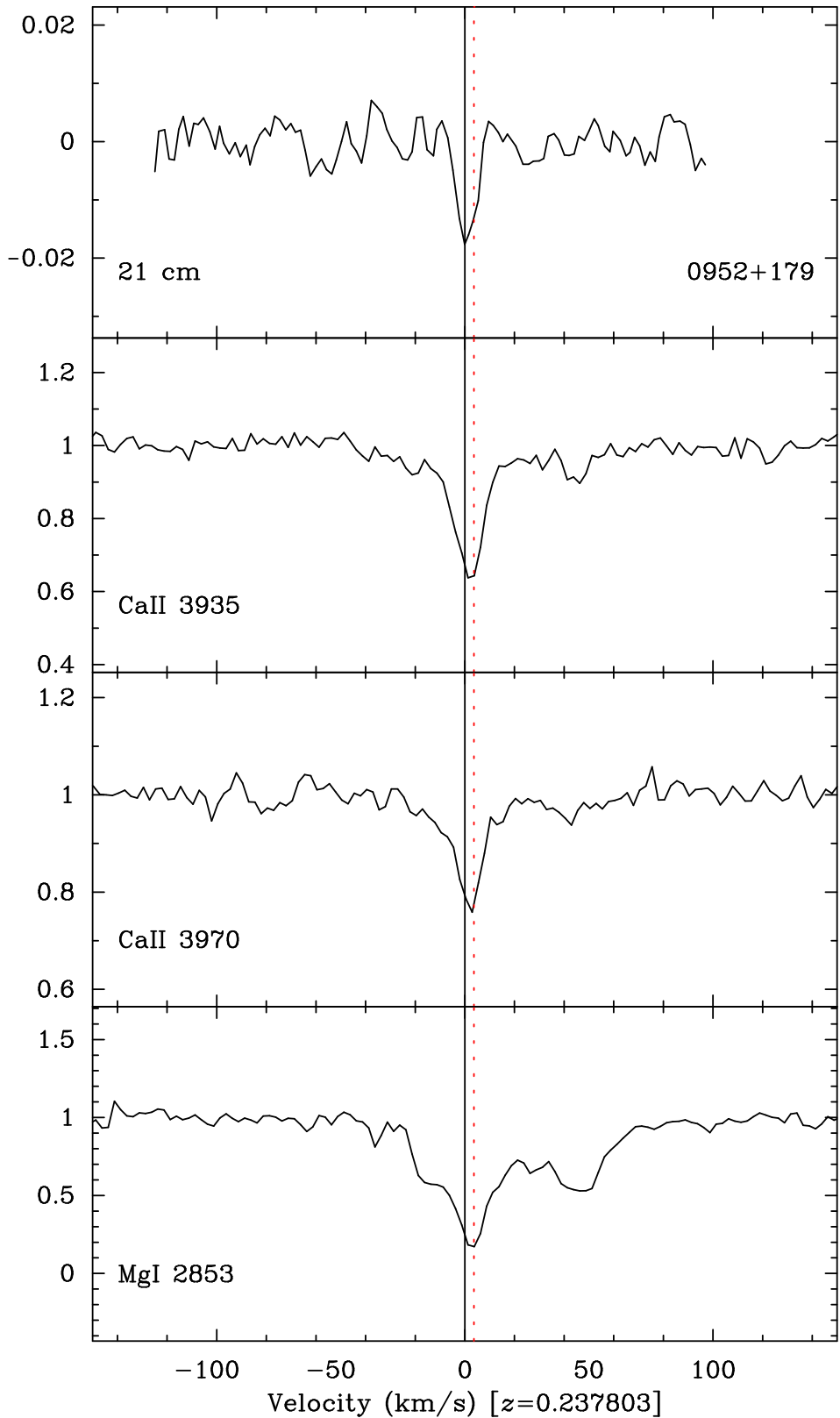}
\caption{\label{fig:0952} Velocity plot for 21-cm and UV absorption
towards quasar \znf.  The 21-cm data are in units of flux (Jy).
The quasar continuum (1.4~Jy) has been subtracted.
The UV data are continuum normalised. The solid vertical line at 0~\kmps\ is at
\zrad. The dotted vertical line is at $\langle z_{\rm UV}\rangle$.
In this and subsequent plots \catwo\ is shown for illustration only but
has not been used in the calculation of the plotted $\langle z_{\rm UV}\rangle$.}
\end{figure*}

\begin{figure*}
\includegraphics[width=1.4\columnwidth, angle=0]{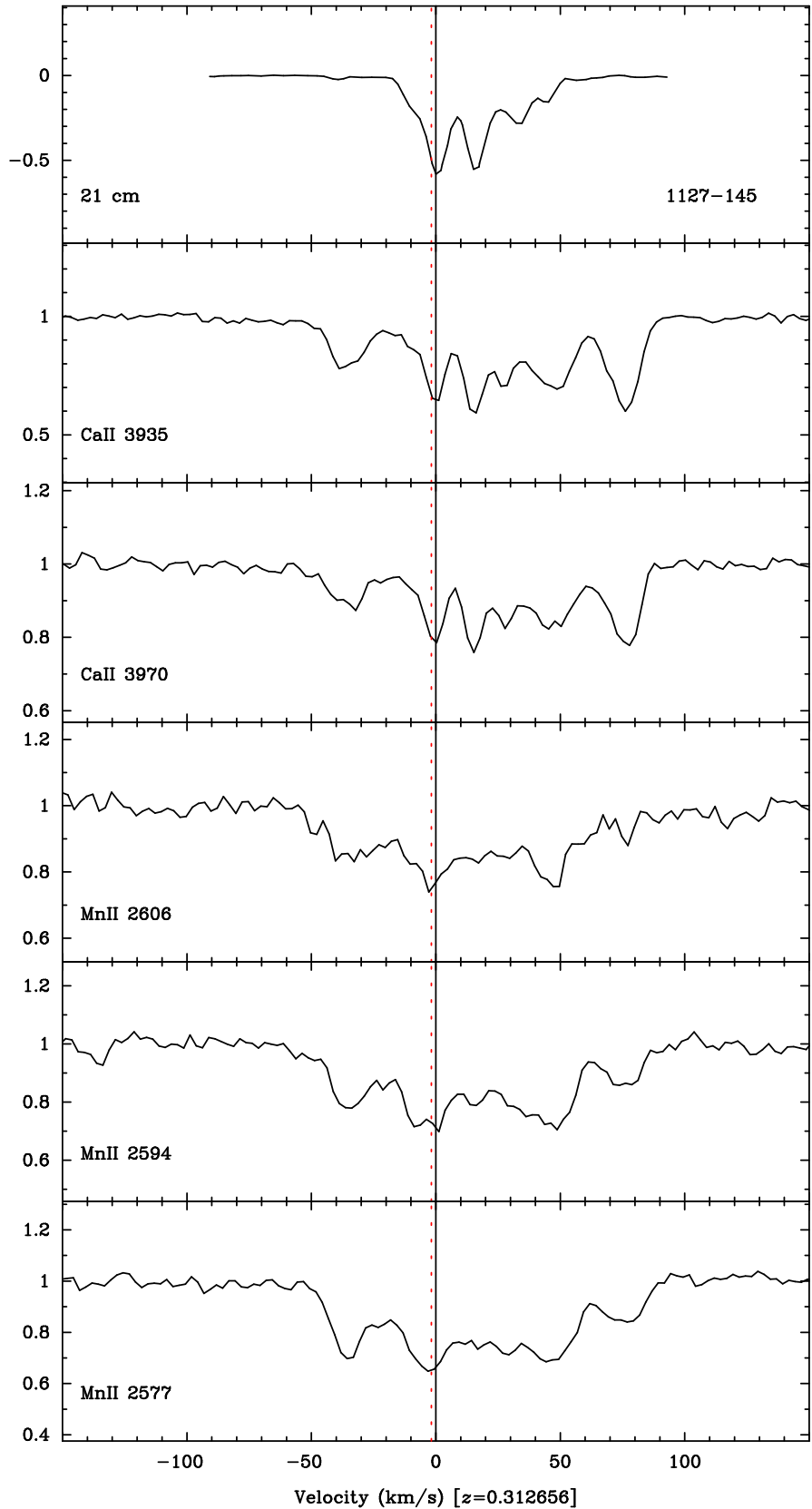}
\caption{\label{fig:1127L} Velocity plot for 21-cm and UV absorption
towards quasar \oot.  The 21-cm data are in units of flux (Jy).
The quasar continuum (6.3~Jy) has been subtracted.
The UV data are continuum normalised. The solid vertical line at 0~\kmps\ is at
\zrad. The dotted vertical line is at $\langle z_{\rm UV}\rangle$.}
\end{figure*}

\begin{figure*}
\includegraphics[width=1.4\columnwidth, angle=0]{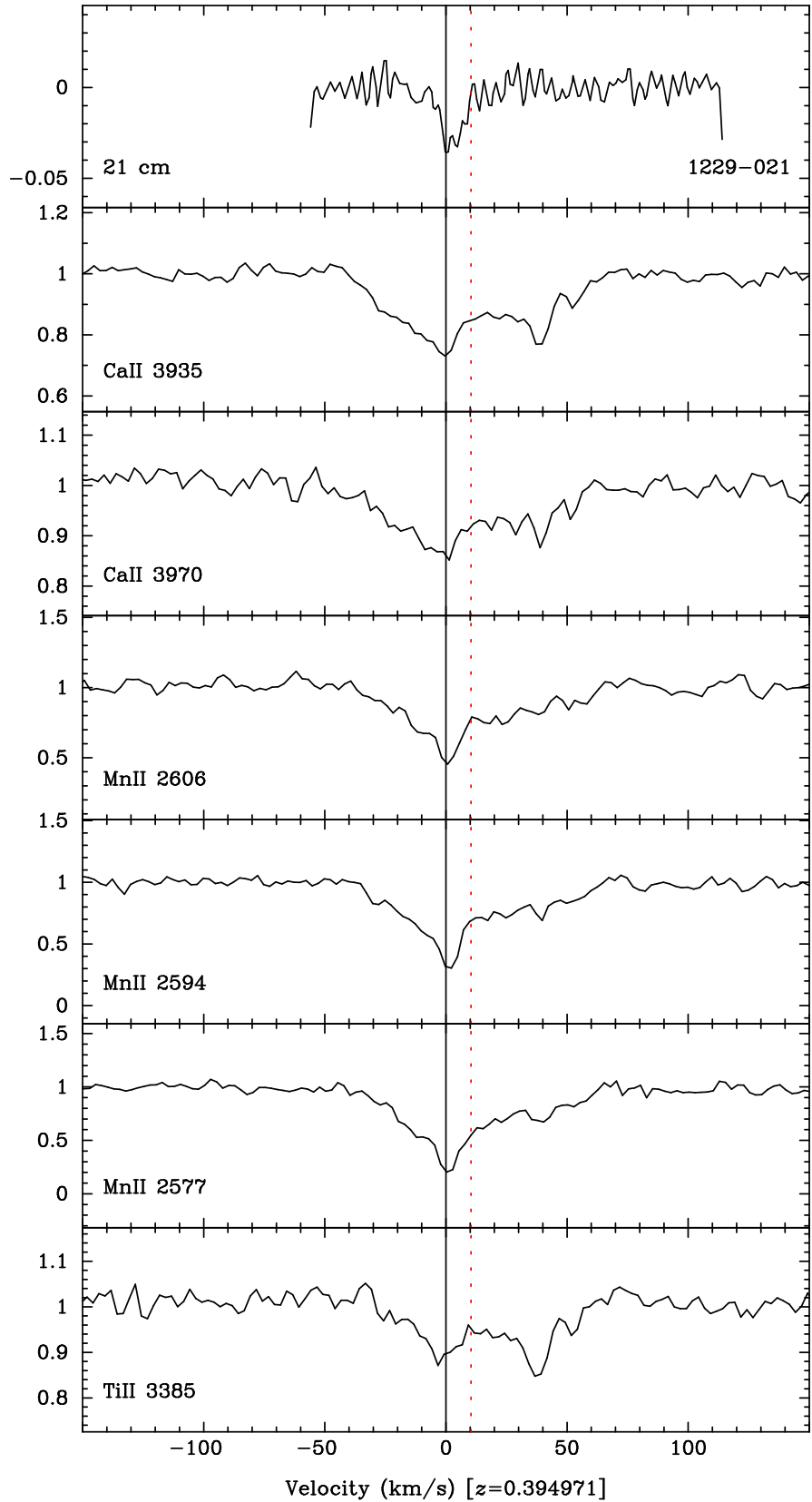}
\caption{\label{fig:1229} Velocity plot for 21-cm and UV absorption
towards quasar \ott.  The 21-cm data are in units of
antenna temperature (K), where the system temperature
was 45~K. The UV data are continuum normalised.
The solid vertical line at 0~\kmps\ is at
\zrad. The dotted vertical line is at $\langle z_{\rm UV}\rangle$.}
\end{figure*}

\begin{figure*}
\includegraphics[width=1.5\columnwidth, angle=0]{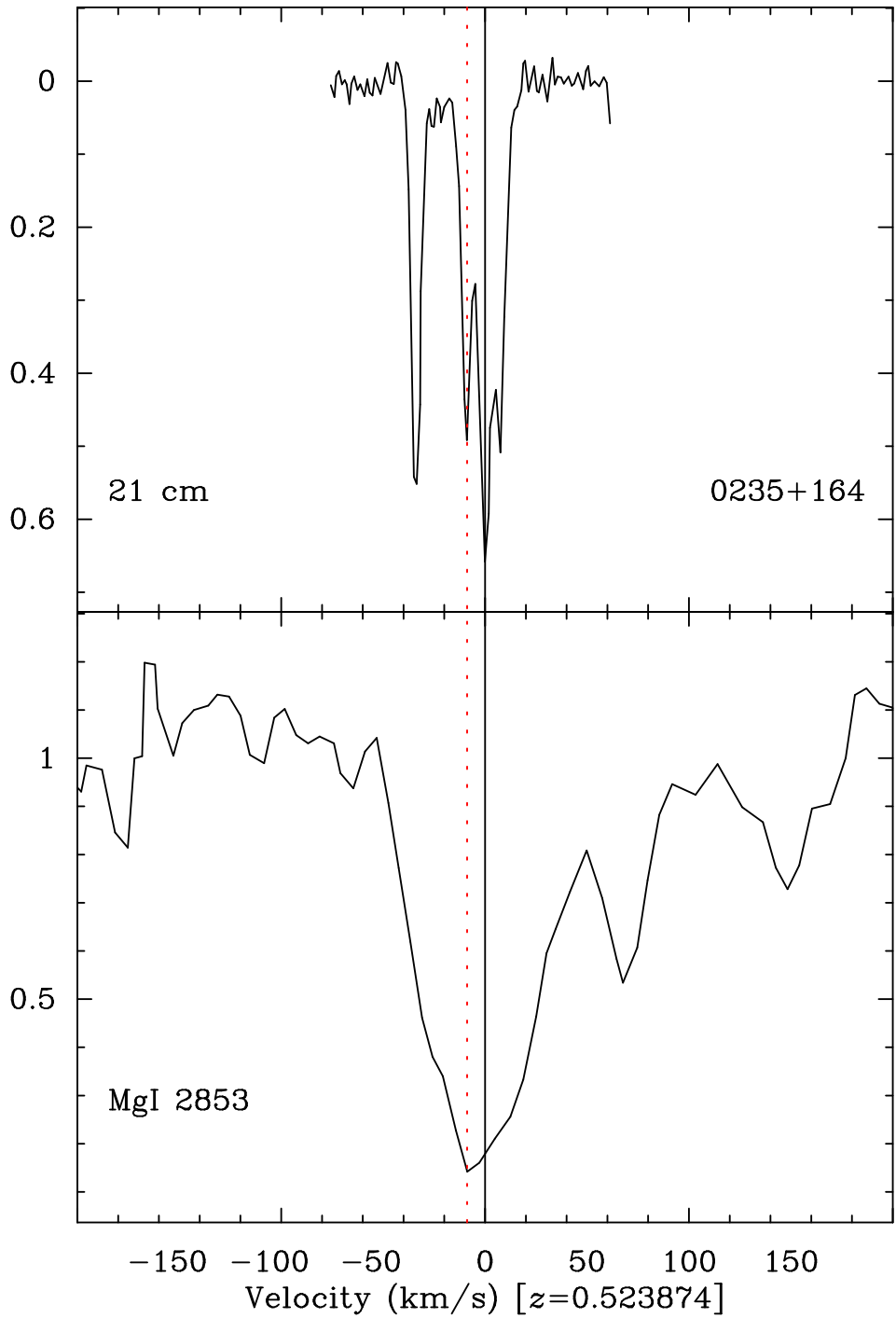}
\caption{\label{fig:0235} Velocity plot for 21-cm and UV absorption
towards quasar \ztt.  The 21-cm data are in units
of optical depth. The UV data are continuum normalised. 
The solid vertical line at 0~\kmps\ is at
\zrad. The dotted vertical line is at $\langle z_{\rm UV}\rangle$.}
\end{figure*}

\begin{figure*}
\includegraphics[width=1.5\columnwidth, angle=0]{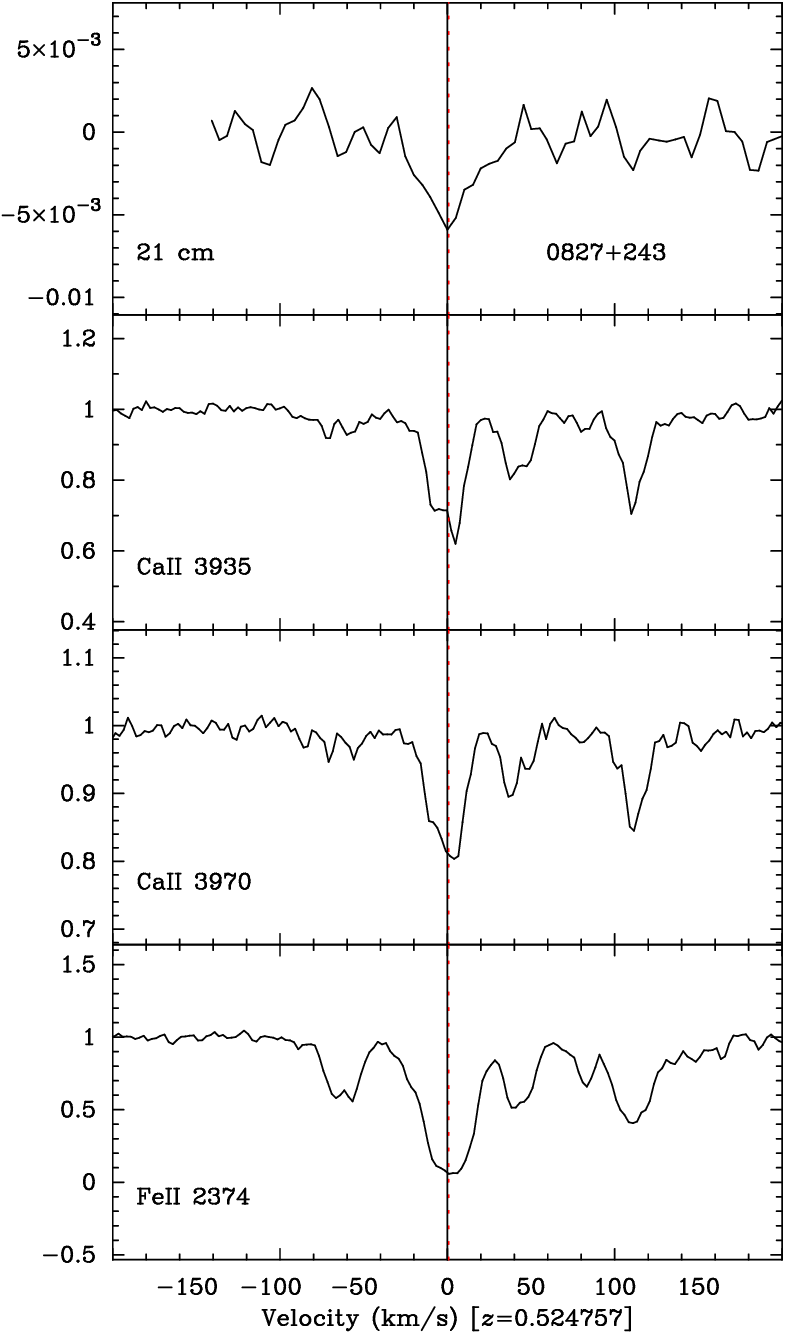}
\caption{\label{fig:0827} Velocity plot for 21-cm and UV absorption
towards quasar \zet.  The 21-cm data are in units of flux (Jy) with
the quasar continuum (0.9~Jy) subtracted.
The UV data are continuum normalised. The solid vertical line at 0~\kmps\ is at
\zrad. The dotted vertical line is at $\langle z_{\rm UV}\rangle$.}
\end{figure*}

\begin{figure*}
\includegraphics[width=1.2\columnwidth, angle=0]{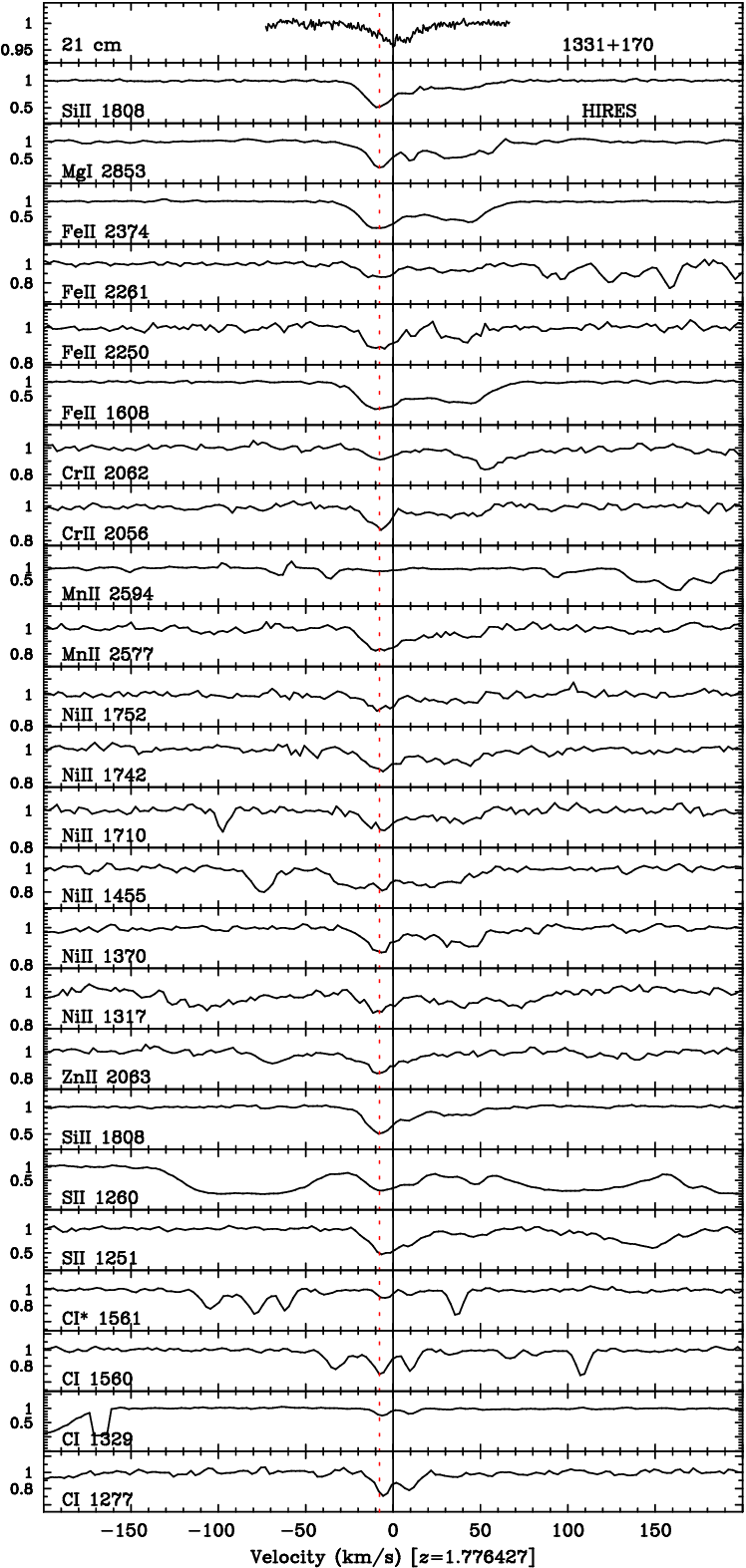}
\caption{\label{fig:1331} Velocity plot for 21-cm and UV absorption
towards quasar \otto.  Both the 21-cm and UV data are continuum
normalised. The solid vertical line at 0~\kmps\ is at
\zrad. The dotted vertical line is at $\langle z_{\rm UV}\rangle$.}
\end{figure*}

\begin{figure*}
\includegraphics[width=1.3\columnwidth, angle=0]{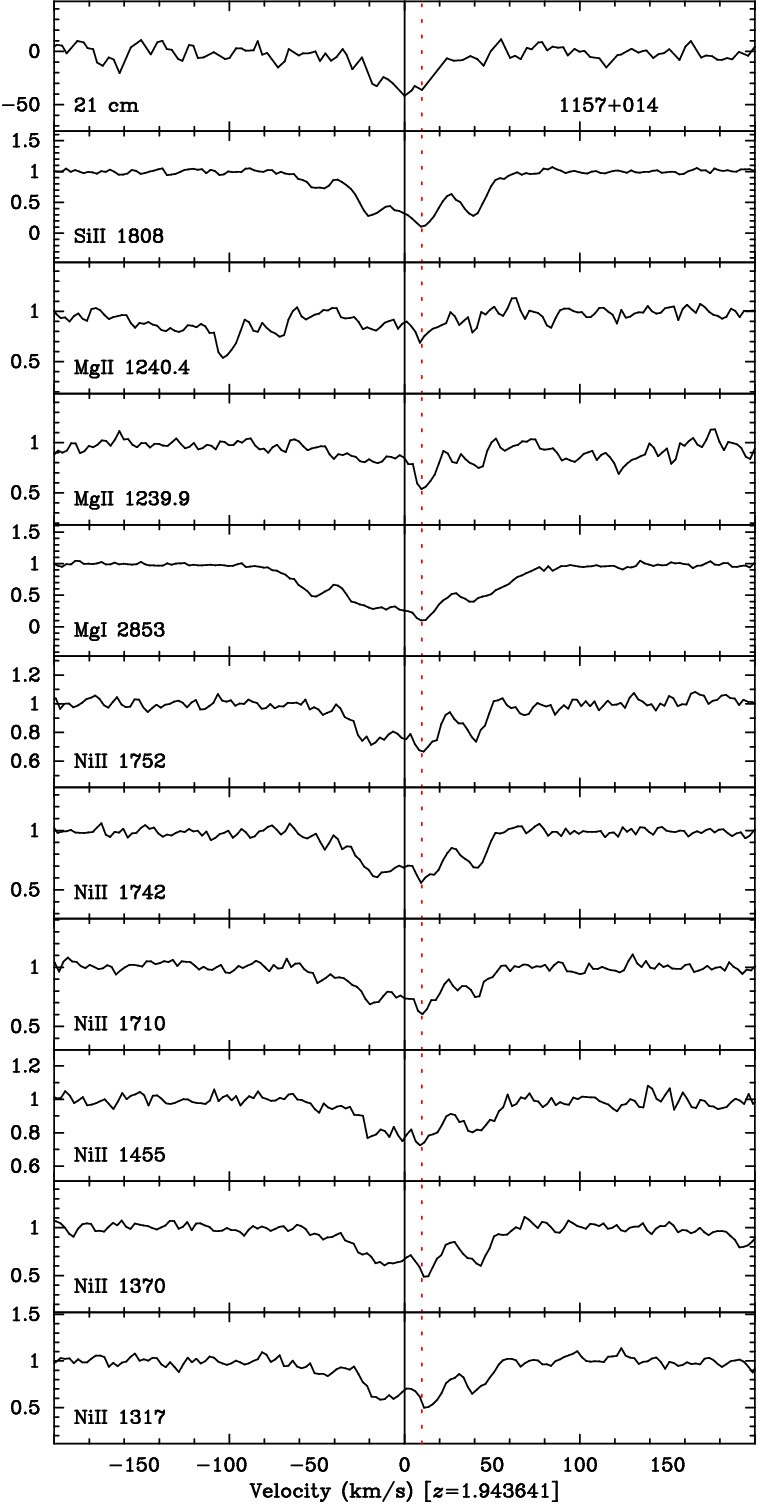}
\caption{\label{fig:1157} Velocity plot for 21-cm and UV absorption
towards quasar \oof.  The 21-cm data are in units of flux density
(mJy) after subtraction of a polynomial baseline. The UV data
are continuum normalised. The solid vertical line at 0~\kmps\ is at
\zrad. The dotted vertical line is at $\langle z_{\rm UV}\rangle$.}
\end{figure*}

\begin{figure*}
\includegraphics[width=1.3\columnwidth, angle=0]{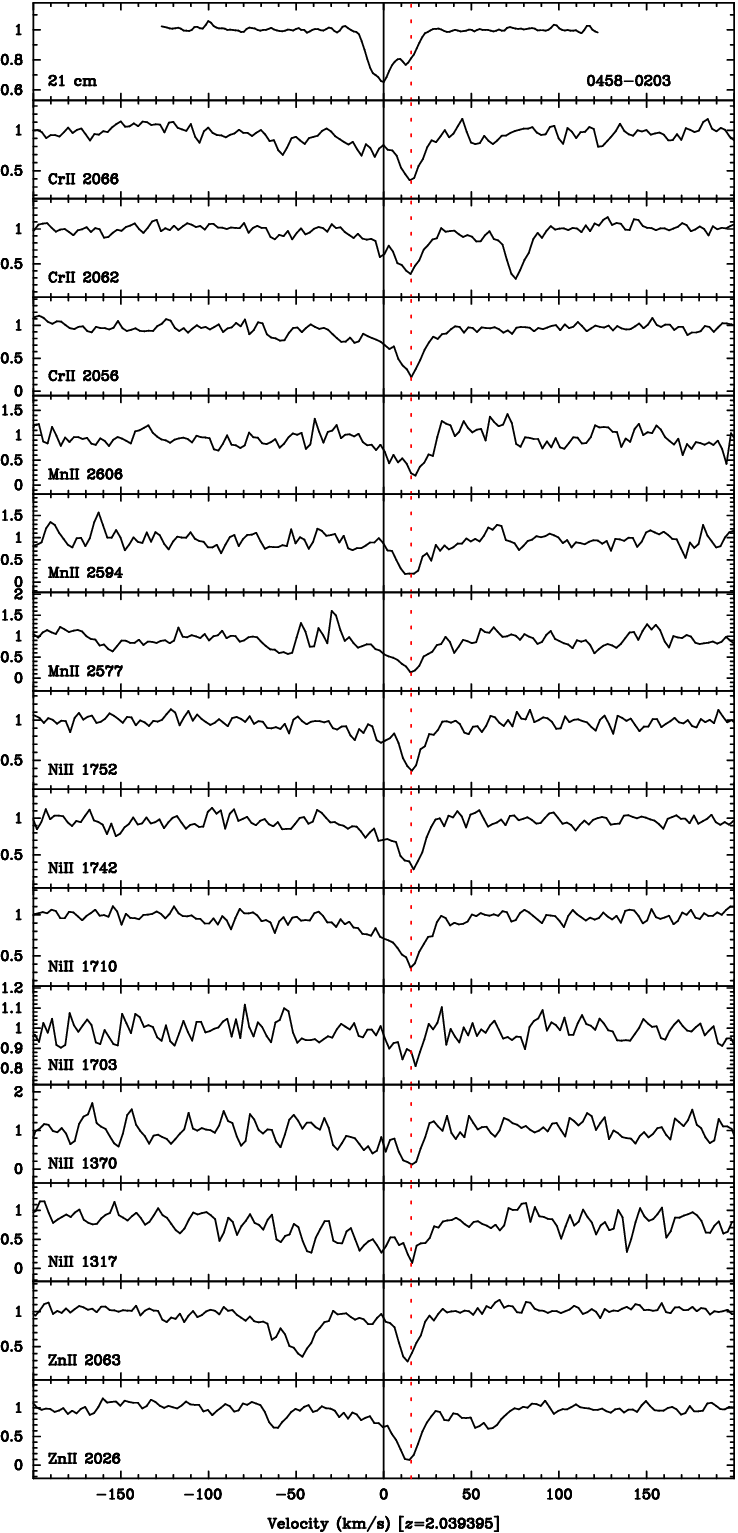}
\caption{\label{fig:0458} Velocity plot for 21-cm and UV absorption
towards quasar \zff.  Both 21-cm and UV data are continuum 
normalised. The solid vertical line at 0~\kmps\ is at
\zrad. The dotted vertical line is at $\langle z_{\rm UV}\rangle$.}
\end{figure*}

\begin{figure*}
\includegraphics[width=1.3\columnwidth, angle=0]{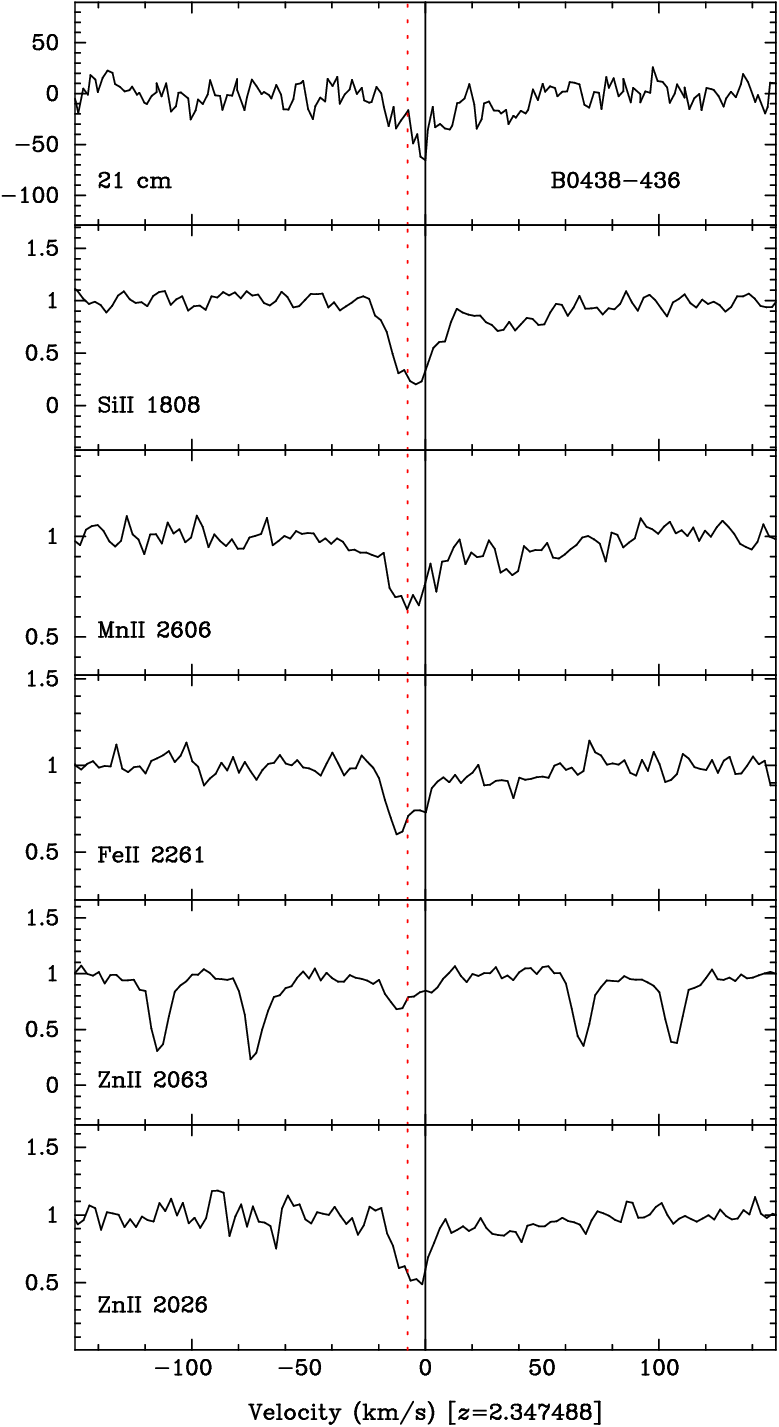}
\caption{\label{fig:0438} Velocity plot for 21-cm and UV absorption
towards quasar \zfe.  The 21-cm data are in units of flux density
(mJy) after subtraction of a polynomial baseline. The UV data are continuum 
normalised. The solid vertical line at 0~\kmps\ is at
\zrad. The dotted vertical line is at $\langle z_{\rm UV}\rangle$.}
\end{figure*}

In the spectral region covered by an absorption system, strongest
absorption occurs at that point where the observed intensity is at a
minimum.  For each 21-cm absorption complex we thus identified and
measured the dispersion coordinate, MHz or \kmps, at the pixel of
minimum intensity, from which we then obtained \zrad. We then searched
the optical data for heavy element absorption features close to the
redshifts where there is 21-cm absorption. We thus identified a number
of UV absorption features, some of which were due to neutral species
and some due to multiply ionised species, with the majority due to
singly ionised species. For all detected UV features, where (1)
absorption was due to either neutral or singly ionised species and (2)
there was little or no saturation, we again determined the value of
the dispersion coordinate, \AA\ or \kmps, at the pixel of minimum
intensity. We then determined the corresponding absorption redshift,
\zuv. When there were several UV transitions in a system, we carried
out this procedure individually for each transition,
e.g.~independently for \zntwo~2026 and \zntwo~2062.  In all, we used
31 different UV transitions. We present detailed velocity plots of all
21-cm and UV absorption components used in this work in
Figs.~\ref{fig:0952}--\ref{fig:0438}. We include the \catwo\ profiles
in these plots for comparison, even though we did not use them to
determine $z_{\rm UV}$. These plots are centred at \zrad, indicated
by the solid vertical line. In each absorption system we calculated
\zavuv, the average of all redshift values determined for single
species, as explained above.  This is shown by the dotted vertical
line in Figs.~\ref{fig:0952}--\ref{fig:0438}.

\subsection{\dxx\ Results}
We have obtained \dxx\ results using three different methods.  The
methods and results are summarised in \tr{tab:res}.  We have also
performed a linear least squares fit to determine the dependence of
$x$ on cosmic time. Additionally, we test evidence for any spatial
variations across the sky.

\subsubsection{Method 1}
For each system we used \zavuv\ together with \zrad\ to obtain
\dxxavq, the average \dxx\ value in a system, from \er{equ:dxx}. Since
each quasar sight-line contains a single detected 21-cm absorber, each
\dxxavq\ value corresponds to a single quasar. These results are
plotted in \fr{fig:dxxz}, where labels are used to indicate which
quasar corresponds to each plotted \dxxavq\ result.

The average of all \dxxavq\ values is $\langle\Delta x/x\rangle_{\rm
  total} = (0.52\pm0.99)\times10^{-5}$.  We label this `{\it result
  1}'.  This covers the absorption redshift range $0.23\approxlt
z_{\rm abs}\approxlt 2.35$, corresponding to look-back times between
2.7 and 10.5 billion years \footnote{We use a Hubble parameter $H_0 =
  73$~\kmps\ Mpc$^{-1}$, a total matter density $\Omega_M=0.27$ and a
  cosmological constant $\Omega_{\Lambda}=0.73$ throughout this paper.
  The age of the Universe in this model is 13.3 billion years.}. For
this result we take as our quoted error the standard deviation on the
mean result, \citep[e.g. see][p.~23]{Meyer1992}, i.e.
\begin{equation}\label{equ:stdmean}
\sigma_{\bar u} = \left[ \frac{\Sigma ( {\bar u}-u_i  )^2 } {n(n-1)} \right]^{0.5} \ ,
\end{equation}
where ${\bar u} \equiv \langle\Delta x/x\rangle_{\rm
total}$, $u_i \equiv $~\dxxavq\ and $n=9$, the number of quasars used.

In other words, for method 1 we have not used individual errors on
\zrad\ and \zavuv\ but instead used the observed scatter to estimate
the error on the mean.  However, as explained next, in methods 2 and 3
we have used the individual errors.  It turns out that these further
calculations provide an {\it a posteriori} justification of method 1
(see \scr{sec:discussion}).

\subsubsection{Method 2}\label{sec:meth2}
As mentioned, for \zrad\ we used the uncertainties given in the
references listed in \tr{tab:data} (see \scr{sec:rad}).  For \zuv\ we
used the standard deviation on \zavuv\ in each absorption system, i.e.
\begin{equation}
\sigma_{\bar w} = \left[ \frac{\Sigma ( {\bar w}-w_i  )^2 } {n(n-1)} \right]^{0.5} \ ,
\end{equation}
where ${\bar w} \equiv \langle z_{\rm UV} \rangle$, $w_i$ is a \zuv\
measurement for a single transition and $n$ is the number of UV
transitions used for an absorption system. We note that it is not
possible to determine $\sigma_{\bar w}$ in the cases of quasars \ztt,
\znf\ and \zet, for which there is only one UV transition (remembering
that we do not use \catwo). Instead, in these cases we have used the
value $6\times 10^{-6}$ which represents a typical error out of all
the \zavuv\ values, except for \ott\ and \zfe\ which are atypical
cases. In the case of \ott, the larger error in \zavuv\ is due to the
fact that the strongest component in \titwo\ 3384 is located at $\sim
+38$~\kmps\ from \zrad. (see \fr{fig:1229}). This is the only case
among the UV transitions used in all spectra where there is such a
large individual discrepancy.  However, for consistency and
simplicity, we have used the redshift value obtained from the \titwo\
3384 component to calculate \zavuv. In the case of \zfe, both the
21-cm spectrum and the UV spectra have quite low signal-to-noise ratio
and dust-depletion effects may give different positions for the
strongest component in different transitions. However, these cases and
the associated larger uncertainties do not reflect the usual situation
for the sample as a whole.

For each absorption system we therefore used \zavuv, \zrad\ and their
uncertainties to calculate the uncertainty in \dxxavq\ (clearly
\dxxavq\ itself remains unchanged for each quasar).  This uncertainty
is shown with the wider error-bar terminators in \fr{fig:dxxz} and
was used to obtain an overall weighted mean result of
\dxxavtw~$=(1.99\pm0.30)\times 10^{-5}$ (`{\it result 2}').

\begin{figure}
\includegraphics[width=1.0\columnwidth]{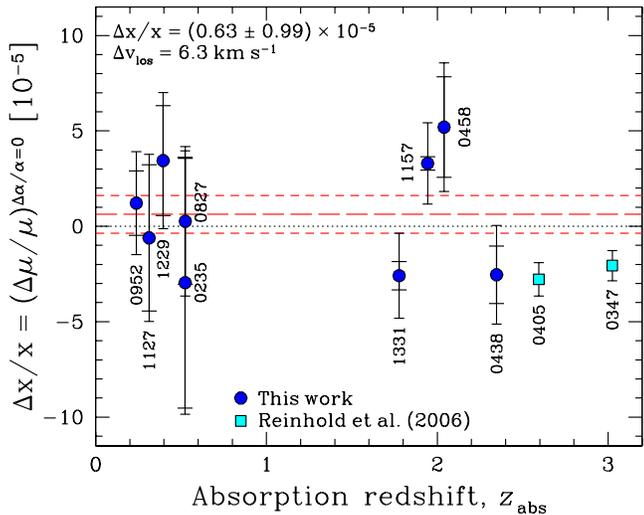}
\caption{\label{fig:dxxz} \dxx\ results for the nine absorption
  systems in our quasar sample. Initial (increased) error bars have
  wider (narrower) error-bar terminators. Quasar names are given
  truncated to four digits. Each point represents \dxxavq\ obtained
  from \zrad\ and $\langle z_{\rm UV}\rangle$, for all heavy element
  species in the UV quasar spectrum, versus average $\langle z_{\rm
    UV}\rangle$ for that spectrum.  The long-dashed horizontal line is
  result 3 and the short-dashed lines show the $\pm 1\sigma$ range.
  Under the assumption of \daa\ $=0$ our results translate to
  constraints on \dmm\ and can be compared with the direct constraints
  from two H$_2$-bearing absorbers in \citet{2006Reinhold}.}
\end{figure}


\subsubsection{Method 3}\label{sec:meth3}
For result 2, the \cs\ per degree of freedom, $\nu$, about the
weighted mean is \csn~$\sim 8$, well beyond what is expected based on
the size of the individual error bars (i.e.~\csn~$\sim 1$). Given the
difficulty in associating 21-cm and UV velocity components, this
provides motivation to quantify this additional scatter by increasing
the individual uncertainties on \dxxavq\ by an amount $s$ so as to
force \csn\ to unity. That is, the total uncertainty for each absorber
is given by
\begin{equation}
  (\sigma_{\langle\Delta x/x\rangle_{\rm abs}}^2 + s^2)^{0.5} .
\end{equation}
A value of \csn~$=1$ is achieved when $s=2.10\times 10^{-5}$. The
error bars now include the statistical uncertainties (as in result 2)
and a systematic component which stems from the problem of comparing
21-cm and UV profiles. With these increased error bars, the weighted
mean result is now \dxxavtw~$= (0.63\pm0.99) \times 10^{-5}$ which we
label `{\it result 3}'. The increased uncertainty for each absorption
system is shown with narrower error-bar terminators in \fr{fig:dxxz}.
We argue below that this new error reflects much better the inherent
uncertainty in the determination of \dxx\ and, as a by-product, it
provides the first reliable estimate for the line-of-sight velocity
differences between 21-cm and UV absorption.  We treat this is our
main result.

\begin{table}
\begin{center}
{\small
\caption{ {\label{tab:res}}\dxx\ result summary. Column 1 gives the label used for
the method and result. Column 2 is a brief description of the error determination
method. Column 3 mentions the type of \dxx\ total result. Column 4 gives
the actual result. Result 3 is our fiducial constraint on \dxx.}
\begin{tabular}{cclc}
\hline
Label & Error type       & \dxx\ type & Result [$10^{-5}$] \\ \hline
1 & Standard error       &$\langle\Delta x/x\rangle_{\rm total}$& $0.52\pm0.99$\\
  & on the mean          &                                   &              \\
2 & Formal statistical   &\dxxavtw\                          & $1.99\pm0.30$\\
  & propagation          &                                   &              \\
3 & Increased error-bars &\dxxavtw\                          & $0.63\pm0.99$\\\hline
\end{tabular}
}
\end{center}
\end{table}

\subsubsection{Variation in $x$ with cosmic time}
Using the increased error bars above, we performed a linear least
squares fit, \dxxavq~$=-\dot{x}/x\,\langle t_{\rm lb}\rangle_{\rm
  abs}$, where $\langle t_{\rm lb}\rangle_{\rm abs}$ is the average
look-back time calculated from the \zavuv\ values in an absorption
system. The best fit gives $\dot{x}/x=(-0.6\pm1.2)\times 10^{-15}{\rm
  \,yr}^{-1}$. Note that by choosing this fitting function, the
intercept of which is zero, we implicitly assume that $\Delta
x/x=0$ at $z=0$. Although this may be corroborated by terrestrial
experiments, it has not been checked in other locations corresponding
to $z=0$ with the same technique used here (e.g.~using 21-cm/UV
absorbers elsewhere in our own galaxy).

\subsubsection{Angular variation in $x$ across the sky}
\begin{figure}
\includegraphics[width=1.0\columnwidth]{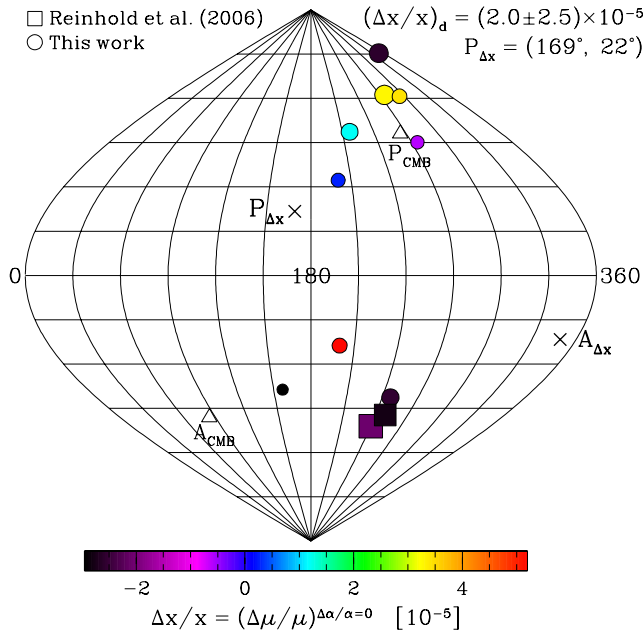}

\caption{\label{fig:aitoff} \dxx\ results for our nine absorbers
  plotted over the sky in Galactic coordinates. The grey- or
  colour-scale encodes the \dxxavq\ values shown in \fr{fig:dxxz} with
  increased error-bars. The size of the data points scale with the
  inverse error-bar. The pole ($P_{\Delta x}$) and anti-pole
  ($A_{\Delta x}$) of the best-fit dipole variation in $x$ is marked
  and the amplitude of the dipole is given. It is not statistically
  significant.  The CMB pole and anti-pole are marked for comparison.
  Assuming \daa\ $=0$, our results constrain \dmm\ and can be compared
  with the more direct constraints on \dmm\ by \citet{2006Reinhold}. 
  Including these additional points, the pole
  moves to $P_{\Delta\mu}=(190^\circ,35^\circ)$ and the amplitude
  becomes $(\Delta\mu/\mu)_{\rm d}=(4.9\pm1.7)\times 10^{-5}$, which
  is formally significant. However, the dipole model is only preferred
  over the constant/monopole model at the 1.6-$\sigma$ level (see
  text).}
\end{figure}


If very large scale ($\sim$10\,Gpc) spatial variations exist, one
might expect $x$ to be different in different directions on the sky.
We plot the distribution of \dxx\ over Galactic coordinates ($l,b$) in
\fr{fig:aitoff}. The grey-scale indicates \dxxavq\ and the size of
each point scales with the inverse of the error on \dxxavq.

There is no obvious dipole (or otherwise) angular variation evident
in \fr{fig:aitoff}. We used a $\chi^2$ minimisation algorithm to find
the best-fit dipole in the angular distribution of \dxx. A grid of
directions in equatorial coordinates, (RA, DEC), was constructed and
for each direction, a cosine fit to \dxx\ as a function of the angular
distance, $\phi$, to each absorption system is performed. This yields
an amplitude for the dipole, $($\dxx$)_{\rm d}$, defined by
\begin{equation}\label{eq:cosine}
  \Delta x/x(\phi) = \left<\Delta x/x\right>^{\rm weighted}_{\rm total} + \left(\Delta x/x\right)_{\rm d}\cos{\phi}\,.
\end{equation}
The best-fit dipole has an amplitude $(\Delta x/x)_{\rm d}=(2.0 \pm
2.5)\times 10^{-5}$ and is in the direction $P_{\Delta x}({\rm
  RA},\rm{\rm DEC})=(7.0{\rm \,hr}, 48^\circ)$ or $P_{\Delta
  x}(l,b)=(169^\circ, 22^\circ)$ as marked on \fr{fig:aitoff}. The
dipole is not significant, despite the fact that we identified the
best-fit direction (see further discussion below).

\subsection{\dmm\ results}
Results on \dxx\ can be used directly to obtain limits on the
variation of $\mu$.  From the definition of $x$ we obtain
\begin{equation}\label{equ:dmm}
  \frac{\mu_z-\mu_0}{\mu_0} \equiv \Delta\mu/\mu =\Delta x/x -
  2\Delta\alpha/\alpha -\Delta g_{\rm p}/g_{\rm p} \ .
\end{equation}
Furthermore, the dependence of $g_{\rm p}$ on fundamental constants is
fairly weak: $\Delta g_{\rm p}/g_{\rm p} \approx -0.1 \frac{\Delta
  (m_q/\Lambda_{\rm QCD})} {(m_q/\Lambda_{\rm QCD})}$
\citep{2004Flambaum}, where $\Lambda_{\rm QCD}$ is the quantum
chromodynamic scale. Therefore the $g_{\rm p}$ term in \er{equ:dmm}
can be neglected, so that there is a simple relation between \dmm,
\dxx\ and \daa. Note that relations between the variation of
$\alpha$, fundamental masses and $\Lambda_{\rm QCD}$
based on Grand Unification Theories are discussed, e.g.,  by
\citet{1984PRL...52...489M,2002EurPhysJC...24...639C,2002PhysLetB...528...121L,
2002PRD...66...045022O,2003NucPhysB...653...256D,2003PRD...67...015009D}
and references therein. 
Within these models the variation of $\alpha$ is $1-2$ orders of magnitude
smaller than the variation of fundamental masses and $\Lambda_{\rm QCD}$,
so it may be neglected in the variation of $x$.
However, these relations between the variations of different fundamental
constants are strongly model-dependent.
 
Combining our result 3 for \dxx\ with the claimed
detection of variations in $\alpha$ from \citet{2003MNRAS.345..609M}
gives $\Delta\mu/\mu=(1.7\pm1.0)\times 10^{-5}$. In this case, the
non-zero \daa\ drives the resulting non-zero \dmm. Using the claimed
null-result on \daa\ of \citet{2004AA...417..853C} yields
$\Delta\mu/\mu=(0.75\pm1.0)\times 10^{-5}$. Although these two
varying-$\alpha$ results are quite different, note that the error-bar
in the resulting value of \dmm\ is completely dominated by the error
in \dxx.  Alternatively, by conservatively assuming \daa~$=0$, we
obtain \dmm\ $=$~\dxxavtw~$= (0.63\pm0.99) \times 10^{-5}$.

Note that the dependence of the proton mass on the current quark mass,
$m_q$, is very weak 
\citep[$\Delta m_p/m_p \approx 0.05 \Delta m_q/m_q$,][]{2004Flambaum}.  
To high accuracy the proton mass is proportional
to the strong interaction parameter $\Lambda_{\rm QCD}$.  Therefore,
the limit on $\mu\equiv m_{\rm e}/m_{\rm p}$ is simultaneously a limit on the more
fundamental Standard Model parameter $\nu\equiv m_{\rm e}/\Lambda_{\rm QCD}$
since $\Delta \mu/\mu \approx \Delta \nu/\nu$.

\subsubsection{Variation in $\mu$ at high-redshift}

\citet{2006Reinhold} used two H$_2$-bearing absorbers, one at
$z_{\rm abs}=2.594$ towards Q0405$-$443 and the other at $z_{\rm
  abs}=3.025$ towards Q0347$-$383, to derive \dmm\ $=(-2.78\pm0.88)$
and $(-2.06\pm0.79)\times 10^{-5}$, respectively, (formally) indicating
a significant detection of cosmological variations in $\mu$. 
Their quoted mean result is $(-2.4\pm0.6)\times 10^{-5}$ (weighted fit)
and $(-2.0\pm0.6)\times 10^{-5}$ (unweighted fit).
Taken at face value, the error-bar on \dmm\ from our results is comparable
with the precision of these latest direct constraints on \dmm.
In that case, even assuming that \daa\ $=0$ at high redshift,
our null result 3 on \dxxavtw\ appears to
contrast
sharply with these more direct measurements of \dmm.

The Reinhold et al. data are plotted in \fr{fig:dxxz} for comparison
with our constraints on \dxx. If the apparent discrepancy between
these two data-sets is to be resolved by appealing to time-variations
in $\mu$, then clearly very sharp and possibly oscillatory variations
would be required. However, our full data set shows
a significant gap in redshift, $\Delta z \sim 1.25$. This naturally
divides our set into a low- and a high-redshift subset. As the
two Reinhold et al. datapoints are at high redshift,
it is clear that they are best compared with our
subset of four high-redshift points, the results for which
are shown in \tr{tab:dmmres}. As can be seen in this table,
apart from the case of pure statistical propagation of errors, our results
have a factor of $\sim 2$ lower precision than those of 
Reinhold et al., and, given the uncertainties, are, in fact, in
agreement ($2\sigma$). 

\begin{table}
\begin{center}
{\small
\caption{ {\label{tab:dmmres}} \dmm\ result summary.
These results were produced in the same way as those tabulated in \tr{tab:res} but
only the four highest-redshift absorbers have been used. Rows 1, 2 and 3a are
for \daa~$=0$,  row 3b for \daa\ from \citet{2003MNRAS.345..609M} and
row 3c for \daa\ from \citet{2004AA...417..853C}.
}
\begin{tabular}{cclc}
\hline
Label & Error type       & \dmm\ type & Result [$10^{-5}$]  \\ \hline
1 & Standard error       &$\langle\Delta \mu/\mu\rangle_{\rm total}$  & $0.84\pm2.00$ $^{1}$\\
  & on the mean          &                                   &              \\
2 & Formal statistical   &\dmmavtw\                          & $2.04\pm0.31$ $^{1}$\\
  & propagation          &                                   &              \\
3a & Increased error-bars &\dmmavtw\                          & $0.58\pm1.95$ $^{1}$\\
3b & &                        & $1.73\pm1.96$ $^{2}$\\
3c &  &                          & $0.70\pm1.95$ $^{3}$ \\

\hline
\end{tabular}
}
\end{center}
1:\daa~$=0$; 2:\daa\ from \citet{2003MNRAS.345..609M}; 3:\daa\ from \citet{2004AA...417..853C}
\end{table}

\subsubsection{Variation in $\mu$ with cosmic time}
Using a similar approach to that for $\dot{x}/x$ above, we assume a
linear variation of $\mu$ with time to constrain $\dot{\mu}/\mu$.
With the increased error bars on the \dxxavq\ values, assuming \daa\
$=0$ and including the results of \cite{2006Reinhold}, the
time-variation of $\mu$ is constrained to be
$\dot{\mu}/\mu=(1.7\pm0.5)\times 10^{-15}{\rm \,yr}^{-1}$. Again, we
assume that \dmm\ $=0$ at zero redshift but, once more, we emphasize
that no absorption-line checks at $z=0$ have been made to justify this
assumption.

\subsubsection{Angular variation in $\mu$ across the sky}

Figure~\ref{fig:aitoff} shows the two Reinhold et al.~absorbers
plotted on the sky with our \dmm\ values (i.e.~our \dxxavq\ values
assuming \daa\ $=0$). Given the concentration of points with \dmm\ $<0$
below the Galactic equator and the mix of values above, it is tempting
to explain the apparent discrepancy between these two data-sets by
spatial variations in $\mu$ across the sky. We performed a similar
dipole search as for $x$ above, including the Reinhold et al.~data,
finding the best-fit dipole direction to be $P_{\Delta\mu}({\rm
  RA},\rm{\rm DEC})=(8.5{\rm \,hr}, 33^\circ)$ or
$P_{\Delta\mu}(l,b)=(190^\circ, 35^\circ)$. The dipole amplitude is
$(\Delta\mu/\mu)_{\rm d}=(4.9\pm1.7)\times 10^{-5}$.

However, the limited range of angular separations in
Fig.~\ref{fig:aitoff} severely limits the dipole interpretation.
Although the dipole amplitude is formally significant at the
2.9-$\sigma$ level, this is only after we had selected the best-fit
direction; this does not answer the question `how significantly is the
dipole model preferred over the constant (i.e.~monopole) model?'. We
address this question using a bootstrap technique. Bootstrap samples
are formed by randomising the values of \dmm\ over the quasar
sight-line directions and the best-fitting dipole direction and
amplitude is determined for each bootstrap sample. The resulting
probability distribution for $(\Delta\mu/\mu)_{\rm d}$ indicates that
values at $>$2.9-$\sigma$ significance occur 12\,per cent of the time
by chance alone. Therefore, the data only marginally support
significant angular variations in \dmm\ given the sparse coverage of
the sky.

\begin{table*}
\begin{center}
  \caption{\small {\label{tab:alt}}
    Alternative results.  The results shown in
    Columns 2, 3 and 4 were obtained by introducing the changes
    briefly described in Column 1 (see text for more details).  All
    results shown here were obtained using method 3.}
\begin{tabular}{lcccc}
\hline
Change & \dxxavtw$/10^{-5}$ & \dmm$/10^{-5}$$^{(a)}$ & \dmm$/10^{-5}$$^{(b)}$ & $\Delta v_{\rm los}$\\\hline
\footnotesize
quoted result            & $0.63\pm0.99$ & $1.78\pm1.02$ & $0.63\pm0.99$ & 6.3~\kmps\ \\
\zrad\ from literature   & $0.97\pm1.05$ & $2.11\pm1.07$ & $0.97\pm1.05$ & 6.8~\kmps\ \\
$+$ no digitised plots   & $1.20\pm1.22$ & $2.35\pm1.24$ & $1.20\pm1.22$ & 8.0~\kmps\ \\
\zrad\ from Gaussians    & $0.49\pm0.94$ & $1.64\pm0.96$ & $0.49\pm0.94$ & 5.8~\kmps\ \\
with \catwo\             & $0.71\pm0.96$ & $2.38\pm0.99$ & $0.71\pm0.96$ & 6.1~\kmps\ \\\hline
\end{tabular}

$^a$\daa\ from \citet{2003MNRAS.345..609M};
$^b$\daa~$=0$.\\
\end{center}
\end{table*}


\section{Discussion}\label{sec:discussion}
The main result of this work is that the data show no evidence for any
time variation of \dxx.  It can be seen from \fr{fig:dxxz} that the
\dxxavq\ points show significant scatter beyond that expected based on
their statistical (i.e.~original) error bars. The error in result 1
directly reflects this scatter. After taking into account individual
redshift errors and propagating them statistically, we need to
increase individual uncertainties so as to obtain a \csn\ $=1$ about
the weighted mean. Even so, this procedure (method 3) gives an overall
error which is essentially the same as for result 1.  This simply says
that {\it the scatter in \dxxavq\ completely dominates any errors in
  individual redshift measurements}. This is an {\it a posteriori}
justification of the validity of method 1, in which individual
redshift errors were neglected and not propagated statistically. The
importance of this has not been realised before, as this is the first
time that \dxx\ has been determined for a sample of objects. If one
determined \dxx\ for a single quasar, there would be no scatter, but
there would still be a hidden and unquantifiable systematic error.  In
practise, of course, for a sample of systems, as we have here, this
additional error has random sign and magnitude and can be treated
statistically.  However, our result shows that any purely statistical
determination of uncertainties, i.e.~without taking the observed
scatter into account, would inevitably underestimate the true
uncertainty.

For example, \citet{1995ApJ...453..596C} used neutral carbon lines
(\cone\ and \cone$^{\ast}$) observed in the Keck/HIRES absorption
spectrum of \otto, to obtain \dxx~$=(0.70\pm0.55_{\rm stat})\times
10^{-5}$. For the same object we obtain \dxx~$=(-2.59\pm0.74_{\rm
  stat}\pm2.10_{\rm syst})\times 10^{-5}$. The similarity in the
statistical errors demonstrates our point. Here our statistical error
is just
\begin{equation}
\left[ \frac{ (\sigma_{\langle z\rangle_{\rm UV}} ^2 + \sigma_{z_{\rm 21}}^2 ) }
       { (1+z_{\rm 21} )^2} \right]^{0.5}  \ ,
\end{equation}
and the systematic error is simply the value for the additional
uncertainty $s$ which gives a \csn\ value of 1 (\scr{sec:meth3}). We
can only determine this because we have several objects. The central
value of \citet{1995ApJ...453..596C} is different because they used a
\zuv\ value which is a weighted mean for observed components
\citep{1994Natur.371...43S} and leads to a value very close to \zrad.
Further explanation of the discrepancy in the central value may come
from the fact that, in this system, we have used 23 distinct UV heavy
element transitions with strongest components well defined within a
few \kmps.

The preponderance of the scatter over redshift uncertainties in
determining the overall error also eliminates any worry about errors
introduced via the digitisation process. We emphasise that having the
original digital data would not have lead to appreciably different
results in this work.

What does this quantified scatter, $s$, mean? Simply from looking at
the 21-cm and UV absorption profiles in
Figs.~\ref{fig:0952}--\ref{fig:0438}, it clearly means that the
assumption that 21-cm and UV absorption occurs at the same physical
location is not correct. In other words the 21-cm and UV absorbing
gases are randomly offset in space. A rough estimate of the average
line-of-sight velocity difference is given by $\Delta v_{\rm los} \sim
cs = 6.3$~\kmps. We may explain this difference qualitatively. The
emitting 21-cm quasar source may appear to have a large angular size
to the absorber. A 21-cm sight-line can then intersect a cold, neutral
hydrogen cloud with little or no heavy elements, whilst a UV/optical
sight-line can intersect another cloud with heavy elements at quite a
different velocity. Such a large angular size may be due to the
combined effects both of physical proximity of the absorber to the
quasar and of intrinsic size of the radio emitting region.  An
alternative explanation involving small scale motion of the
interstellar medium has been suggested by \citet{2000PhRvL..85.5511C}
in the case of the $\sim 10$~\kmps\ velocity difference between 21-cm
and mm absorption lines.

The robustness of our \dxx\ and \dmm\ results, as well as our estimate
of $\Delta v_{\rm los}$, can be investigated by means of alternative
calculations. \tr{tab:alt} summarises the simple tests we have carried
out.  All results quoted here have been obtained by applying method 3.
Firstly, if we use \zrad\ directly from the literature, we obtain, for
\dxxavtw, \dmm\ \citep[with \daa\ from][]{2003MNRAS.345..609M} and \dmm\
(with \daa~$=0$), $0.97\pm1.05$, $2.11\pm1.07$ and $0.97\pm1.05$,
respectively, and in units of $10^{-5}$
(row 2 in \tr{tab:alt}). If, additionally, we exclude any results for
\ztt\ and \oot, we obtain $1.20\pm1.22$, $2.35\pm1.24$ and $1.20\pm1.22$
(row 3 in \tr{tab:alt}). We have also fitted Gaussians to the 21-cm
absorption profiles and used these to calculate \zrad, i.e. we used
the location of the minimum in the fitted Gaussian profile rather than
the pixel of minimum intensity. We then obtained $0.49\pm0.94$,
$1.64\pm0.96$ and $0.49\pm0.94$ (row 4 in \tr{tab:alt}). Finally, we
have investigated what happens if one includes the \catwo\ lines in
the calculations as well. This gives $0.71\pm0.96$, $2.38\pm0.99$ and
$0.71\pm0.96$.  Note that, because we include \catwo, we have no
problem calculating $\sigma_{\bar w}$ (\scr{sec:meth2}) for quasars
\ztt, \znf\ and \zet.

It is thus obvious that our results are relatively insensitive to the
method used to estimate absorption redshifts, and, perhaps, even to
the ionisation level of the transitions used. The dominant factor is
the scatter due to the velocity mismatch between 21-cm and UV
transitions.  Because the uncertainties in our determinations of UV
redshifts are small when compared to this velocity mismatch, we do not
expect that standard Voigt profile fitting of the UV absorption
features would significantly affect our results either. The mean
line-of-sight velocity differences may become comparable in
magnitude to individual errors in redshift measurements only if a much
larger ($\sim 100$) sample of 21 cm/UV absorbers is obtained: The average
uncertainty per absorption system due to the velocity offset is
now $\sim 10^{-5}$ but with 100 absorbers, or so, we would expect this to
be reduced by a factor of $\sqrt {100}$, i.e. to $\sim 10^{-6}$. 
A definitive answer to the question of whether \dxx\ varies, or not,
with cosmic time, will have to wait until then.

Another important result is that, even when we conservatively assume
\daa\ $=0$, our measurements of \dxx\ imply constraints on \dmm\ which
are independent of the current
possible detections of \citet{2006Reinhold}. 
These
authors find a combined (unweighted) value of
$\Delta\mu/\mu=(-2.0\pm0.6)\times 10^{-5}$ from two H$_2$-bearing
absorbers while we obtain a less direct constraint of
$\Delta\mu/\mu=(0.58\pm1.95)\times 10^{-5}$.
This new constraint,
based on four high-redshift 21-cm/UV absorbers combined 
with the assumption of \daa\
$=\Delta g_{\rm p}/g_{\rm p}=0$, serves as an internally robust,
independent comparison to the H$_2$ measurements.

\section*{Acknowledgements}
We thank the anonymous referee for
constructive comments which helped improve this paper, 
Ralph Spencer for providing useful observation information
with regards to the radio spectrum of \ott\ and Art Wolfe for
providing original data and for useful discussions. MTM thanks PPARC
for the award of an Advanced Fellowship. This work was partially funded by the
Australian Research Council, the Particle Physics and Astronomy
Research Council (UK) and the Department of Energy, Office of Nuclear
Physics, under Contract No. W-31-109-ENG-38 (USA).  Some of the
observations have been carried out using the VLT at the Paranal
observatory.  Some of the data presented herein were obtained at the
W.~M. Keck Observatory, which is operated as a scientific partnership
among the California Institute of Technology, the University of
California, and the National Aeronautics and Space Administration. The
Observatory was made possible by the generous financial support of the
W.~M. Keck Foundation.

\label{lastpage}

\newcommand{\noopsort}[1]{}

\vspace{-2cm}

\vspace{-2cm}

\vspace{-2cm}

\vspace{-2cm}

\end{document}